\documentclass[useAMS,usenatbib]{mn2e}

\usepackage{graphicx,amssymb}

\newcommand{\plotwd}{8.5cm}
\newcommand{\plotwdtwo}{15cm}

\title[Power spectrum of the maxBCG sample]{Power spectrum of the maxBCG sample: detection of acoustic oscillations using galaxy clusters}
\author[Gert H\"utsi]{Gert H\"utsi$^{1,2}$\thanks{E-mail: ghutsi@star.ucl.ac.uk}\\
$^{1}$Department of Physics and Astronomy, University College London, London, WC1E 6BT\\
$^{2}$Tartu Observatory, EE-61602 T\~oravere, Estonia}
\begin{document}

\date{}

\pagerange{\pageref{firstpage}--\pageref{lastpage}} \pubyear{}

\maketitle

\label{firstpage}

\begin{abstract}
We use the direct Fourier method to calculate the redshift-space power spectrum of the maxBCG cluster catalog \citep{2007astro.ph..1265K} -- currently by far the largest existing galaxy cluster sample. The total number of clusters used in our analysis is $12,616$. After accounting for the radial smearing effect caused by photometric redshift errors and also introducing a simple treatment for the nonlinear effects, we show that currently favored low matter density ``concordance'' $\Lambda$CDM cosmology provides a very good fit to the estimated power. Thanks to the large volume ($\sim 0.4 \,h^{-3}\,\rm{Gpc}^3$), high clustering amplitude (linear effective bias parameter $b_{\rm{eff}}\sim 3\times(0.85/\sigma_8)$), and sufficiently high sampling density ($\sim 3 \times 10^{-5} \,h^{3}\,\rm{Mpc}^{-3}$) the recovered power spectrum has high enough signal to noise to allow us to find weak evidence ($\sim 2 \sigma$ CL) for the baryonic acoustic oscillations (BAO). In case the clusters are additionally weighted by their richness the resulting power spectrum has slightly higher large-scale amplitude and smaller damping on small scales. As a result the confidence level for the BAO detection is somewhat increased: $\sim 2.5\sigma$. The ability to detect BAO with relatively small number of clusters is encouraging in the light of several proposed large cluster surveys.
\end{abstract}


\section{Introduction}
Since the flight of the {\sc Cobe} satellite in the early 1990's the field of observational cosmology has witnessed extremely rapid progress which has culminated in the establishment of the Standard Model for cosmology -- the ``concordance'' model \citep{1999Sci...284.1481B,2003ApJS..148..175S}. This progress has been largely driven by the precise measurements of the angular temperature fluctuations of the Cosmic Microwave Background (CMB) \citep{2000ApJ...545L...5H,2002ApJ...571..604N,2003ApJS..148....1B,2003MNRAS.341L..23G,2003ApJ...591..556P}. However, in order to break several degeneracies inherent in the CMB measurements one has to complement this data with other sources of information, such as the measurements of the SNe~Ia luminosity distances \citep{1998AJ....116.1009R,1999ApJ...517..565P} or with the measurements of the large-scale structure (LSS) of the Universe as traced by galaxies or galaxy clusters. 

The simplest descriptor one can extract from the LSS measurements is the matter power spectrum. The broad-band shape of this spectrum is sensitive to the shape parameter $\Gamma = \Omega_mh$, and thus is useful in helping to establish the currently favored low matter density ``concordance'' model, as well as the amount of baryons in relation to the total matter $f_b=\Omega_b/\Omega_m$. Currently the two largest galaxy redshift surveys are the 2dF Galaxy Redshift Survey\footnote{http://www.mso.anu.edu.au/2dFGRS/} (2dFGRS) and the Sloan Digital Sky Survey\footnote{http://www.sdss.org/} (SDSS) with its latest data release 5 (DR5), providing redshifts to $\sim 220,000$ and $\sim 675,000$ galaxies, respectively. The SDSS galaxy sample consists of two broad classes: (i) the MAIN galaxy sample, reaching redshifts of $z \sim 0.25$; (ii) the Luminous Red Galaxy (LRG) sample covering redshift range $z \sim 0.15 - 0.5$. Some other characteristics of these surveys are listed in Table \ref{tab}.  

\begin{table*}
\caption{Some characteristics of the 2dFGRS, SDSS DR5, and maxBCG samples.}
\label{tab}
\begin{tabular}{l|l|l|l|l|l}
Survey & Number of & Sky area & Redshift & Comoving volume & Effective bias parameter wrt to\\
 & objects & (deg$^2$) & coverage & ($h^{-3}\,\rm{Gpc}^3$) & the model with $\sigma_8=0.85$\\
\hline
\hline
2dFGRS & $\sim 220,000$ & $\sim 1200$ & $\lesssim 0.3$ & $\sim 0.07$ & $\sim 1.0$ \\
\hline
SDSS DR5 & $\sim 675,000$ & $\sim 5700$ & MAIN: $\lesssim 0.25$ & MAIN: $\sim 0.2$ & MAIN: $\sim 1.2$\\
& & & LRG: $\sim 0.15 - 0.5$ & LRG: $\sim 1.3$ & LRG: $\sim 2.0$\\
\hline
maxBCG & $\sim 13,800$ & $\sim 7000$ & $ 0.1 - 0.3$ & $\sim 0.4$ & $\sim 3$
\end{tabular}
\end{table*}

In addition to the precise measurement of the broad-band shape of the power spectrum, SDSS LRG and 2dFGRS samples have proven useful in allowing the detection of theoretically predicted oscillatory features in the spectrum. The source of these spectral fluctuations is the same as that giving rise to the prominent peak structure in the CMB angular power spectrum -- acoustic oscillations in the tightly coupled baryon-photon fluid prior to the epoch of recombination \citep{1970Ap&SS...7....3S,1970ApJ...162..815P}. However, the relative level of fluctuations in the matter power spectrum is strongly reduced compared to the corresponding fluctuations in the CMB angular spectrum, owing to the fact that most of the matter in the Universe is provided by the cold dark matter (CDM) that does not participate in acoustic oscillations. After recombination, or more precisely after the end of the so-called ``drag-epoch'' which happens at redshifts of a few hundred, as the baryons are released from the supporting pressure of the photon gas they start to fall back to the CDM potential wells that have started to grow already since the matter-radiation equality. Since the amount of baryons in the total matter budget is not completely negligible -- baryons make up roughly one fifth -- the acoustic structure imprinted onto the baryonic component also gets transferred to the spatial distribution of the CDM. After baryonic and CDM components have relaxed one ends up with $\sim 5\%$ relative fluctuations in the total mater power spectrum.

In order to observe these relatively small fluctuations one needs to have:
\begin{enumerate}
\item Large survey volumes, to reduce cosmic variance;
\item High enough spatial sampling density, to decrease discreteness noise.
\end{enumerate}
Evidently, for the fixed telescope time there is a trade-off between these two requirements. Amongst the available spectroscopic samples the optimum is currently best met by the SDSS LRGs, and indeed, at the beginning of 2005 the detection of the acoustic peak in the spatial two-point correlation function was announced by \citet{2005ApJ...633..560E}. At the same time the final power spectrum measurements of the 2dFGRS also revealed the existence of the acoustic features \citep{2005MNRAS.362..505C}. Baryonic acoustic oscillations (BAO) in the power spectrum of the SDSS LRG (DR4) sample were found by \citet{2006A&A...449..891H}\footnote{For an independent comparison see \citet{2008PThPh.120..609Y}.} and the corresponding cosmological parameter estimation was carried out in \citet{2006A&A...459..375H}. Also, several more recent detections of BAO have been reported: \citet{2006astro.ph..5302P} and \citet{2007MNRAS.374.1527B} used SDSS photometric LRG sample, \citet{2006PhRvD..74l3507T,2008ApJ...676..889O,2009MNRAS.393.1183C,2008arXiv0807.3551G,2009ApJ...696L..93M,2009arXiv0901.2570S,2009arXiv0908.2598K} analyzed spectroscopic LRG sample, \citet{2007ApJ...657..645P} carried out a combined analysis of the SDSS MAIN and LRG spectroscopic galaxy samples, and finally, \citet{2007MNRAS.381.1053P,2009arXiv0907.1660P} analyzed SDSS LRG, MAIN and 2dFGRS samples. In the recent paper by \citet{2009PhRvD..79f3512N} a weak detection of the BAO damping is reported.

The usefulness of BAO arises from the fact that it can provide us with a standard ruler in the form of the size of the sound horizon at decoupling\footnote{To be more precise again, at the end of the ``drag-epoch''. In the ``concordance'' $\Lambda$CDM model the sound horizon at the ``drag-epoch'' is $\sim 5\%$ larger compared to the one at recombination.}, enabling us to carry out a purely geometric cosmological test: comparing the apparent size of the ruler along and perpendicular to the line of sight with the physical size of the sound horizon (which can be well calibrated from the CMB data) one is able to find Hubble parameter $H(z)$ and angular diameter distance $d_A(z)$ corresponding to the redshift $z$ (e.g. \citealt{2003PhRvD..68f3004H}). Obviously, having determined $H(z)$ and $d_A(z)$ one can put constraints on the dark energy (DE) equation of state parameter $w_{DE}(z)$. In that respect radial modes, which enable us to find $H(z)$, are more useful since $H(z)$ is related to $w_{DE}(z)$ through a single integration whereas $d_A(z)$, which is given by the angular modes, involves double integration over $w_{DE}(z)$. Thus one would certainly benefit from accurate redshift information.
  
If one wishes to use the BAO signal as a precise standard ruler there are a few complications one has to overcome:
\begin{enumerate}
\item As the density field goes nonlinear couplings between Fourier modes (and thus the resulting transfer of power) will modify the BAO signal, leading to the damping of the oscillations;
\item In order to fully exploit the information in galaxy/cluster surveys one has to understand the relation of these objects to the underlying matter density field, i.e. one has to have a realistic model for biasing. 
\end{enumerate}
These two complications can be fully addressed only through costly N-body simulations. There are several papers that have investigated the detectability and possible systematics of BAO extraction using the numerical simulations e.g. \citet{1999MNRAS.304..851M,2005Natur.435..629S,2005ApJ...633..575S,2007APh....26..351H,2007A&A...462....7K,2007astro.ph..2543A,2007astro.ph..3620S}. However, significantly more work towards these directions is probably needed before we can harness the full power provided by the potentially very clean geometric tool in the form of BAO in the matter power spectrum.

It is worth pointing out that actually one of the first possible hints for the existence of BAO in the matter power spectrum came from the analysis of the spatial clustering of the Abell/ACO clusters \citep{2001ApJ...555...68M}.\footnote{Interesting features in the spatial two-point correlation function of the Abell/ACO cluster sample were also revealed by \citet{1997MNRAS.289..801E}.} The power spectrum measurement in \citet{2001ApJ...555...68M} used $637$ clusters with richness $R \geq 1$. In this paper we analyze the maxBCG cluster sample \citep{2007astro.ph..1265K} which is currently by far the largest cluster catalog available, spanning a redshift range $z=0.1-0.3$, containing $13,823$ objects, and covering $\sim 0.4 \,h^{-3}\,\rm{Gpc}^3$ of comoving volume (see Table \ref{tab} for comparison with 2dFGRS and SDSS DR5). The biggest advantage of using galaxy clusters instead of the typical, i.e. $L_*$-galaxies, is the fact that their spatial clustering signal is strongly amplified with respect to the clustering of the underlying matter: bias parameter $b$ often reaching values $\sim 3$ and above.
\begin{figure}
\centering
\includegraphics[width=\plotwd]{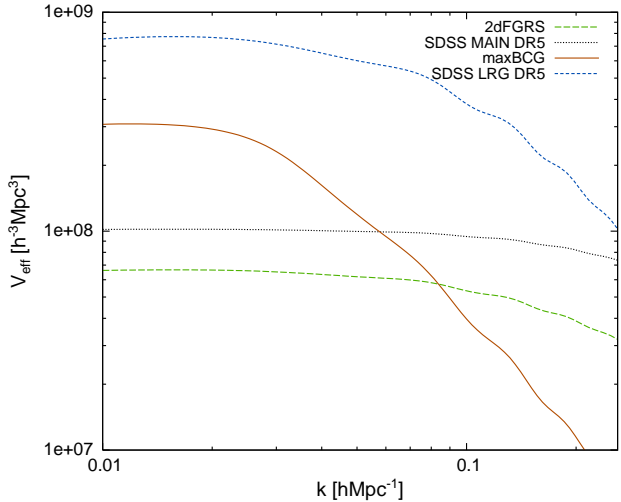}
\caption{The effective volume (see Eq. (\ref{eq2})) of the maxBCG cluster sample in comparison to the 2dFGRS and SDSS MAIN and LRG samples from data release five (DR5).}
\label{fig1}
\end{figure}
In Fig. \ref{fig1} we have compared the effective volume (see Eq. (\ref{eq2})) of the maxBCG survey with the two other surveys that have lead to the detection of BAO: 2dFGRS and SDSS LRG. In addition we have also shown the effective volume of the SDSS MAIN galaxy sample. Under the assumption of Gaussianity the power spectrum measurement errors $\Delta P$ are simply found as: $\Delta P/P = 2/V_{\rm{eff}}/V_k$, where $V_k$ is the volume of the shell in $k$-space over which the angular average is taken, i.e. $V_k = 4 \pi k^2 \Delta k/(2\pi)^3$. It is no surprise that the SDSS LRG sample is currently unbeatable. However, we notice that maxBCG sample should provide better measurement of the power spectrum on scales larger than $k \sim 0.1 \,h\,\rm{Mpc}^{-1}$ compared to the 2dFGRS. The rapid fall of the effective volume of the maxBCG sample on smaller scales is caused by the photometric redshift errors of $\delta z \sim 0.01$ as estimated by \citet{2007astro.ph..1265K}.

Thus one would expect the maxBCG sample to reveal acoustic features. As our further analysis shows this indeed turns out to be the case, albeit with a relatively modest confidence level.

Our paper is organized as follows. In Section 2 we give a brief description of the maxBCG catalog with the corresponding selection effects. The core part of this paper, which is devoted to the power spectrum analysis, is given in Section 3. In Section 4 we test the stability of our spectrum measurements by relaxing several assumptions made in the original analysis. Finally, Section 5 contains our conclusions.

\section{Data and survey selection function}\label{sec2}
A maxBCG cluster catalog\footnote{http://umsdss.physics.lsa.umich.edu/catalogs/\\maxbcg\textunderscore public\textunderscore catalog.dat} \citep{2007astro.ph..1265K} is compiled via the ``red sequence'' cluster detection method \citep{1998AJ....116.2644O,2000AJ....120.2148G,2003ApJS..148..243B} applied to the SDSS photometric data. The catalog contains $13,823$ galaxy clusters with velocity dispersions greater than $\sim 400$ km/s and covers $\sim 7000$ square degrees of sky between redshifts $0.1$ and $0.3$. The photometric redshifts are estimated using the tight relation between the ridgeline color and redshift and are claimed to have accuracy $\delta z \equiv \sqrt{\left < (z_{\rm photo} -z_{\rm spec})^2\right >} \sim 0.01$ essentially independent of redshift. For more details about the catalog we refer to the original source \citet{2007astro.ph..1265K}.

\begin{figure}
\centering
\includegraphics[width=\plotwd]{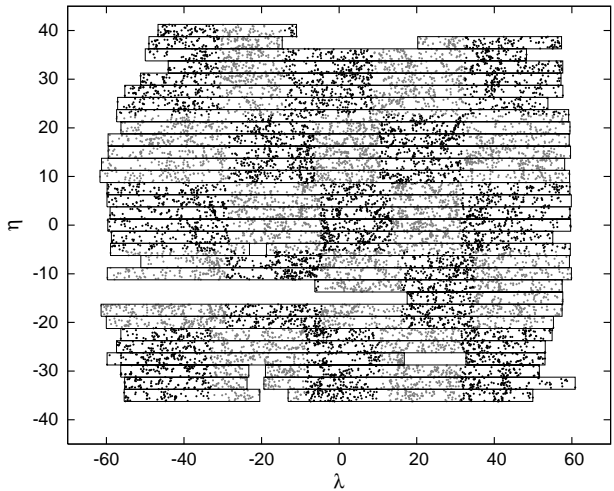}
\caption{Angular distribution of $12,616$ maxBCG clusters in SDSS survey coordinates $(\eta,\lambda)$. The $5 \times 5$ chessboard pattern of black/gray points shows the angular part of the division used for ``jackknife'' error analysis. The union of solid rectangles represents our reconstruction of the angular mask.}
\label{fig2}
\end{figure}

In our power spectrum analysis we are going to neglect the three southern SDSS stripes, leaving us with $12,616$ galaxy clusters over $\sim 6800$ square degrees of sky. With the redshift range $0.1 < z < 0.3$ this corresponds to $\sim 0.4 \,h^{-3}\,\rm{Gpc}^3$ of comoving cosmic volume. Neglecting the three narrow southern stripes helps us to achieve a better behaved survey window function with reduced sidelobes, and thus with reduced leakage of power. 

The angular distribution of the remaining clusters is given in Fig. \ref{fig2}. Here the angular coordinates are plotted using the SDSS survey coordinates $(\eta,\lambda)$ (e.g. \citealt{2002AJ....123..485S}). The chessboard pattern of black/gray points in Fig. \ref{fig2} represents the division of the sky into $5 \times 5$ angular regions used for ``jackknife'' error analysis. We also use three divisions along the redshift direction, making the total of $3\times 5 \times 5 =75$ regions, each containing approximately $168$ clusters. The union of solid rectangles in Fig. \ref{fig2} represents our reconstruction of the survey angular mask. All the rectangles are aligned with the SDSS imaging scan stripes. As the number density of galaxy clusters in the maxBCG sample is rather high, one can determine relatively accurately the beginning, ending, and also possible gaps in the scan stripes. We call the angular mask obtained this way a ``minimal'' mask in contrast to the ``maximal'' mask, which is built in the same manner except that each of the rectangles is extended by an amount corresponding to the mean cluster separation. We carry out our power spectrum analysis using both of these angular masks.   

\begin{figure}
\centering
\includegraphics[width=\plotwd]{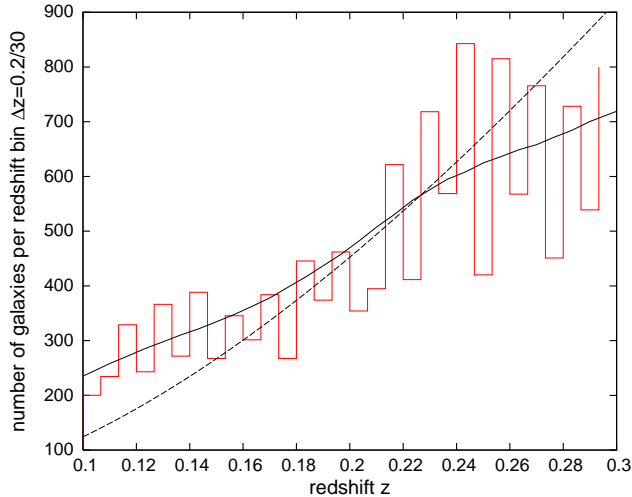}
\caption{Redshift histogram of the maxBCG sample. The solid smooth line corresponds to the cubic spline fit and the dashed line represents the expected number of clusters per redshift bin for the volume-limited sample.}
\label{fig3}
\end{figure}

In Fig. \ref{fig3} we show the redshift distribution of maxBCG clusters. Here the solid smooth line corresponds to the cubic spline fit and the dashed line represents the expected number of clusters per redshift bin $\Delta z = 0.2/30$ for the volume-limited sample\footnote{In \citet{2007astro.ph..1265K} it is suggested that maxBCG sample is rather close to being volume-limited.} assuming a spatially flat $\Lambda$CDM cosmology with $\Omega_m=0.27$. To convert this redshift distribution to the radial selection function $\bar{n}(r)$, i.e. the comoving number density of clusters at distance $r$, we fix the background cosmology the low matter density model mentioned above. Our power spectrum calculations are done for both of the radial selection models shown in Fig. \ref{fig3}.   

For the full survey selection function $\bar{n}(\mathbf{r})$ we assume, as usual, that it can be factorized to the product of angular and radial parts, i.e. $\bar{n}(\mathbf{r})=\bar{n}(\mathbf{\hat{r}})\bar{n}(r)$ ($r\equiv |\mathbf{r}|$, $\mathbf{\hat{r}}\equiv \mathbf{r}/r$). Here $\bar{n}(r)$ is the radial selection function and the angular selection part $\bar{n}(\mathbf{\hat{r}})$ is assumed to take value $1$ if the unit direction vector $\mathbf{\hat{r}}$ is inside the survey mask and $0$ otherwise, i.e. we assume a simple binary angular mask.

\section{Power spectrum analysis}
We calculate the redshift-space power spectrum of the maxBCG catalog using the direct Fourier method as described in \citet{1994ApJ...426...23F} (FKP). Strictly speaking, power spectra determined in this way are so-called pseudospectra, meaning that the estimates derived are convolved with a survey window. Since in the case of the analyzed maxBCG sample the volume covered is very large, reaching $0.4 \,h^{-3}\,\rm{Gpc}^3$, and also the survey volume has relatively large dimensions along all perpendicular directions, the correlations in the Fourier space caused by the survey window are rather compact. On intermediate scales and in the case the power spectrum binning is chosen wide enough, FKP estimator gives a good approximation to the true underlying power.

Instead of direct summation as presented in \citet{1994ApJ...426...23F} we speed up the calculations using Fast Fourier Transforms (FFTs). This gives rise to some extra complications. As our density field, which is built using the Triangular Shaped Cloud (TSC) \citep{1988csup.book.....H} mass assignment scheme, is now ``restricted to live'' on a regular grid with a finite cell size, we have to correct for the smoothing effect this has caused. Also, if our underlying density field contains spatial modes with higher frequency than our grid's Nyquist frequency, $k_{\rm{Ny}}$, then these will be ``folded back'' into the frequency interval the grid can support, increasing power close to $k_{\rm{Ny}}$ -- the so-called aliasing effect. Often there is no need to correct for the aliasing effect since one usually chooses fine enough grid that has Nyquist frequency significantly higher than the highest spatial frequencies of interest. Although the aliasing has a very mild effect on our results we correct it by using the iterative scheme as described in \citet{2005ApJ...620..559J} with a slight modification: we do not approximate the small-scale spectrum by a simple power law, but also allow for the possible running of the spectral index, i.e. the parametric shape of the power spectrum is taken to be a parabola in log--log. We have compared our antialiasing scheme with the original recipe of \citet{2005ApJ...620..559J} finding a full agreement between the two. Thus the measured spectrum along with the acoustic features is very insensitive to the variations in the antialiasing method applied. 

For the full details of our power spectrum calculation method along with several tests we refer the reader to \citet{2006A&A...446...43H,2006A&A...449..891H}. As a very brief summary, our spectrum determination consists of the following steps:
\begin{itemize}
\item Determination of the survey selection function, i.e. mean expected number density without any clustering $\bar{n}(\mathbf{r})$ (see Section \ref{sec2}). This smooth field, with respect to which the fluctuations are measured, is modeled using a random catalog that contains $100$ times more objects than the real maxBCG catalog, i.e. $1,261,600$ in total;
\item Calculation of the overdensity field on a $512^3$-grid using TSC mass assignment scheme;
\item FFT of the gridded overdensity field;
\item Calculation of the raw 3D power spectrum by taking the modulus squared from the output of the previous step;
\item Subtraction of the shot noise component from the raw 3D spectrum;
\item Recovery of the angle averaged spectrum using a modified version of the iterative scheme of \citet{2005ApJ...620..559J}.    
\end{itemize}

\begin{figure}
\centering
\includegraphics[width=\plotwd]{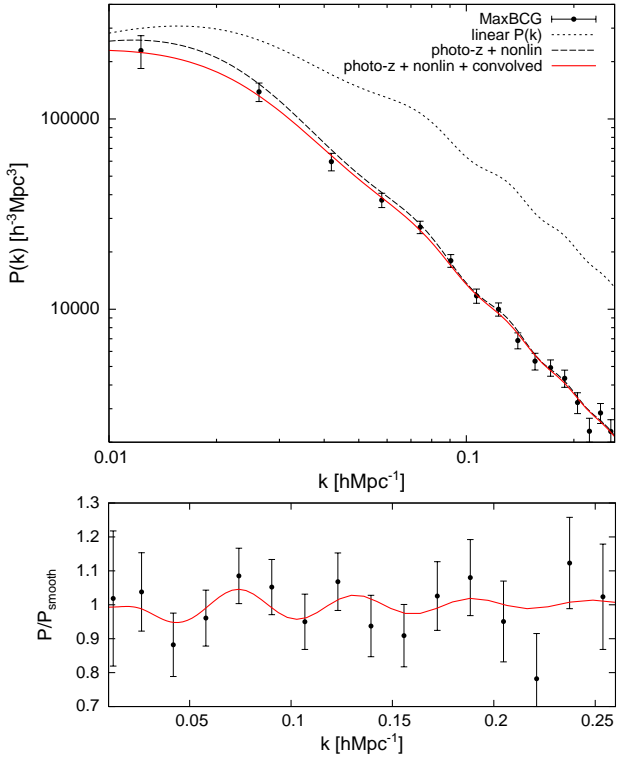}
\caption{Upper panel: Filled circles with errorbars show the redshift-space power spectrum of maxBCG cluster sample. Solid line represents the best fitting model in case of three free parameters: $b_{\rm{eff}}$, $q$, $\sigma$, and assumes Monte Carlo error model. Long dashed line shows the same model without survey window convolution applied and short dashed line corresponds to the linear model spectrum. Lower panel: the same as above, with the broad-band shape of the spectrum removed by dividing with a ``smooth'' model spectrum without BAO.}
\label{fig4}
\end{figure}

\begin{figure}
\centering
\includegraphics[width=\plotwd]{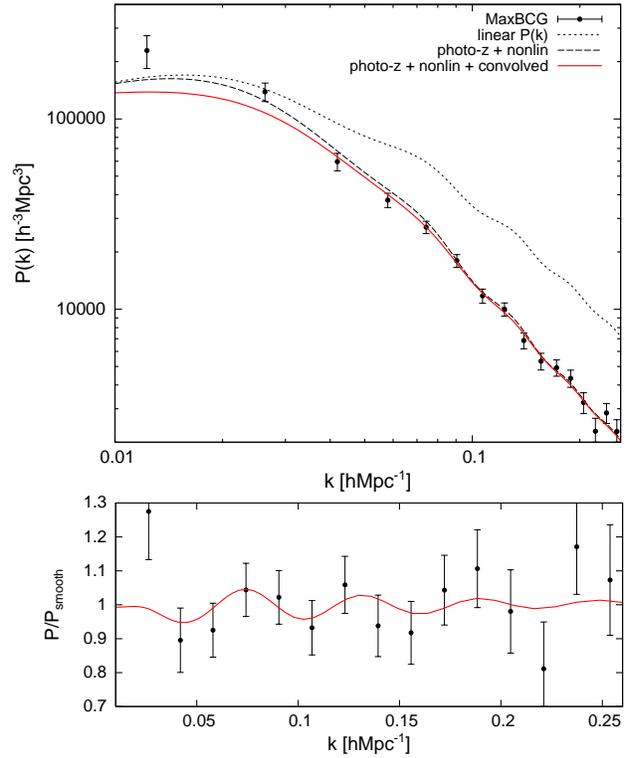}
\caption{Same as Fig. \ref{fig4}, with the only difference that here we have two free parameters: $b_{\rm{eff}}$, $q$, and the value for $\sigma$ is fixed to $30\,h^{-1}\,\rm{Mpc}$.}
\label{fig5}
\end{figure}

The results of our power spectrum calculation along with fitted models are shown in Figs. \ref{fig4} and \ref{fig5}. Here the errorbars $\Delta P$ are calculated as described in FKP, which assumes that the density field follows Gaussian statistics (see also \citealt{1998ApJ...499..555T}):
\begin{equation}
\frac{\Delta P}{P} = \sqrt{\frac{2}{V_{\rm{eff}}V_k}}\,,
\end{equation}
where the effective volume $V_{\rm{eff}}$ (see Fig. \ref{fig1}) is given by:
\begin{equation}\label{eq2}
V_{\rm{eff}} = \Omega\cdot\int\limits_{0}^{1}\,{\rm d}\mu\int\limits_{r_{\min}}^{r_{\max}}\left[ \frac{\bar{n}(r)P(k,\mu)}{1+\bar{n}(r)P(k,\mu)}\right]^2r^2\,{\rm d}r\,, 
\end{equation}
and $V_k$ is the volume of the $k$-space shell over which the angular average of the 3D spectrum is taken, i.e.
\begin{equation}
V_k=\frac{4\pi k^2 \Delta k}{(2\pi)^3}\,.
\end{equation}
In Eq. (\ref{eq2}) $r_{\min}$ and $r_{\max}$ are comoving distances corresponding to the redshift bounds of the survey ($0.1$ and $0.3$ in case of maxBCG), $\bar{n}(r)$ is the radial selection function, and $\Omega$ is the area of the sky in steradians covered by the survey. $P(k,\mu)$ is the anisotropic power spectrum\footnote{Here we assume a flat sky approximation.}  where $\mu$ is the cosine of the angle between the line of sight direction $\hat{\mathbf{r}}$ and wavevector $\mathbf{k}$, i.e. $\mu \equiv \hat{\mathbf{r}}\cdot \hat{\mathbf{k}}$. To calculate effective volumes for the spectroscopic surveys, such as 2dFGRS and SDSS shown in Fig. \ref{fig1}, one can neglect mild distortions due to redshift-space distortions and take power spectrum to be a function of $k$ only, i.e. drop integral over $\mu$ in Eq. (\ref{eq2}). However, for photometric surveys with large distortions along the line of sight this approximation is no longer valid. For the maxBCG survey we fit a parametric model in the form of Eqs. (\ref{eq6}) and (\ref{eq7}) to the measured isotropized power spectrum. To calculate effective volume plotted in Fig. \ref{fig1} we have taken $P(k,\mu)$ as in Eq. (\ref{eq7}) with the replacement $f(k) \rightarrow f(k,\mu)\equiv \exp[-(\mu k \sigma)^2]$. Here $\sigma$ is the spatial smoothing scale along the line of sight: $\sigma = \frac{c}{H_0}\delta z\,$.

In addition to the FKP errors we have also estimated the power spectrum variance via the ``jackknife'' method (see e.g. \citealt{1993stp..book.....L}). For this purpose we have divided the survey volume into $5 \times 5$ angular regions (see Fig. \ref{fig2}) along with $3$ redshift bins, making the total of $75$ regions, each containing on average $168$ clusters. Now the power spectrum is calculated $75$ times with each time one of the regions omitted. The covariance of the power spectrum can now be estimated via the scatter of the power spectrum measurements: 
\begin{eqnarray}
\rm{cov}(P_i,P_j) & = & \frac{N-1}{N}\cdot\sum\limits_{n=1}^{N}\left( P_i^{(n)}-\bar{P_i}\right)\left( P_j^{(n)}-\bar{P_j}\right)\,,\\
\bar{P_i} & = & \frac{1}{N}\cdot \sum\limits_{n=1}^{N}P_i^{(n)}\,.
\end{eqnarray}
Here $P_i^{(n)}$ represents the power spectrum measurement for the $i$-th power spectrum bin in case the $n$-th survey subvolume was excluded from the calculations. 
\begin{figure}
\centering
\includegraphics[width=\plotwd]{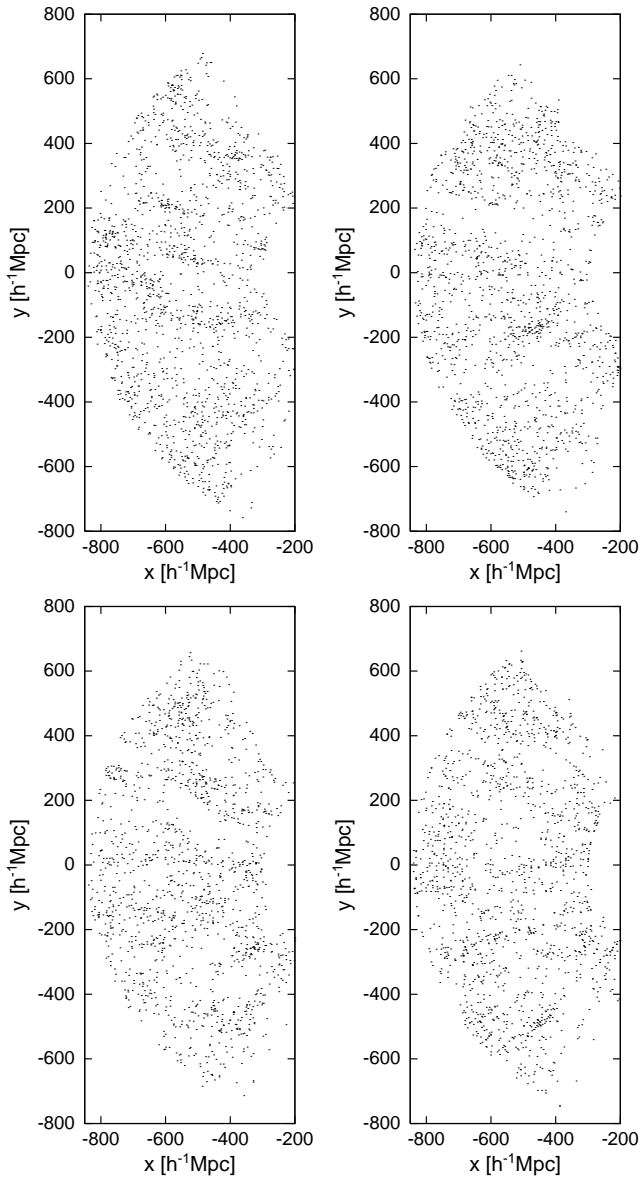}
\caption{Example slices (5 degrees wide) through the mock catalogs and maxBCG sample (bottom right-hand panel). All of the point sets have very similar two-point functions.}
\label{newfig1}
\end{figure}
\begin{figure}
\centering
\includegraphics[width=\plotwd]{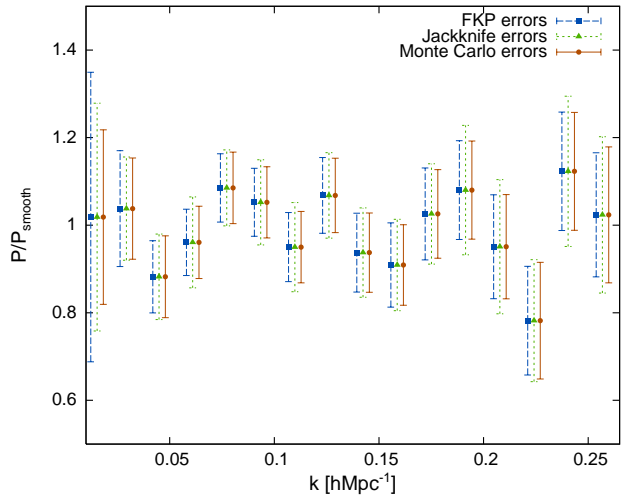}
\caption{Comparison of the FKP, ``jackknife'', and Monte Carlo error estimates. Note that the errors as given here were calculated by dividing each mock spectrum by a smooth fit to it, i.e. the additional uncertainty due to the variation of the power spectrum amplitude is taken out.}
\label{fig7}
\end{figure}
\begin{figure}
\centering
\includegraphics[width=\plotwd]{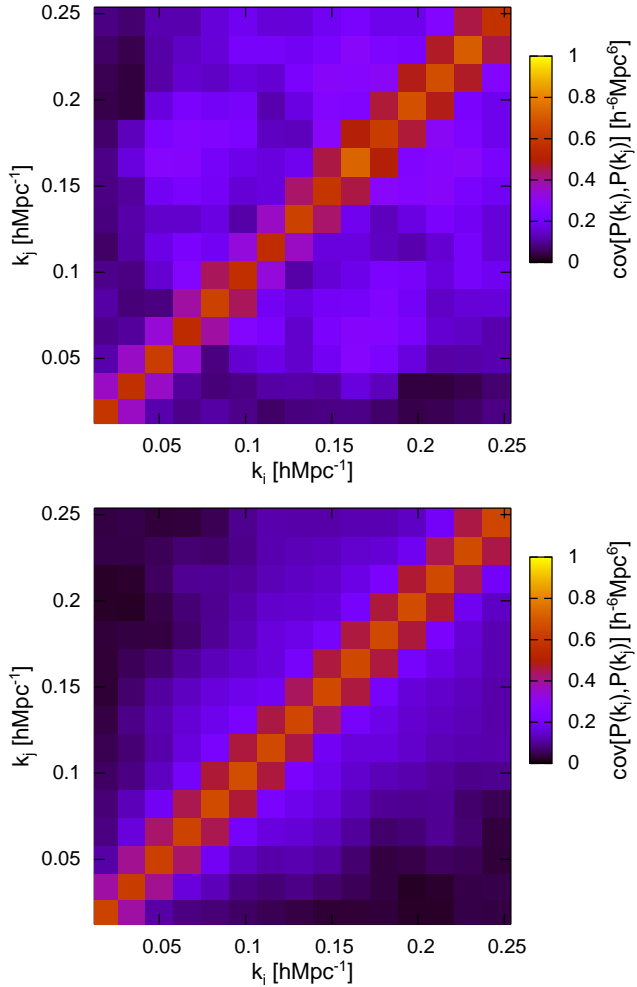}
\caption{``Jackknife'' (upper panel) and Monte Carlo (lower panel) correlation matrices $\rm{corr}(P_i,P_j)\equiv\rm{cov}(P_i,P_j)/\sqrt{\rm{var}(P_i)\rm{var}(P_j)}$.}
\label{newfig2}
\end{figure}
There are several obvious problems related to the ``jackknife'' procedure, e.g. how to choose the size for the subvolumes, also the number of subvolumes in our case is relatively small, leading to a rather noisy covariance matrix. To overcome those shortcomings we have used a third method to estimate the errors. In this Monte Carlo approach we have generated $1000$ mock catalogs with the survey selection function matching that of the maxBCG sample using a method described in detail in Appendix B of \citet{2006A&A...449..891H}. Our method consists of three parts: (i) generation of the density field using the optimized 2nd order Lagrangian perturbation scheme (2LPT), (ii) Poisson sampling of the density field with the sampler whose parameters are tuned to result in the clustering strength similar to the maxBCG sample, (iii) addition of the redshift-space distortions and photo-z errors. In Fig. \ref{newfig1} we show a few five degrees wide slices through mock catalogs along with the slice through the maxBCG sample (bottom right-hand panel). All of the point samples in Fig. \ref{newfig1} have very similar two-point functions and since the 2LPT is able to capture the morphology of the quasilinear density field down to scales of $\sim 10 h^{-1}\,\rm{Mpc}$ we expect also a rather good match for the higher order correlations on those scales.
All of the three error estimates are compared in Fig. \ref{fig7} where for clarity we have removed the smooth broad-band component of the spectrum. We see that in general errors are in good agreement. In particular, the Monte Carlo and FKP errors agree remarkably well on almost all, except for the very largest scales. The mild discrepancy at the largest scales is probably due to the finite box size (which leads to a poor sampling of the very largest Fourier modes) used in our 2LPT simulations. In Fig. \ref{newfig2} we compare the ``jackknife'' and Monte Carlo correlation matrices $\rm{corr}(P_i,P_j)\equiv\rm{cov}(P_i,P_j)/\sqrt{\rm{var}(P_i)\rm{var}(P_j)}$. Due to relatively small number of subvolumes (75) used in our ``jackknife'' analysis the resulting correlation matrix is quite noisy. Also, one can notice that on the largest scales the matrix elements neighboring the main diagonal drop off quite a bit faster in comparison to the Monte Carlo case. In the following model fitting part of our paper we use all three error estimates obtained above. For the case of the FKP errors a simple diagonal covariance matrix is assumed. If not stated explicitly, all of the figures in this paper assume Monte Carlo covariance matrix.   

\begin{figure}
\centering
\includegraphics[width=\plotwd]{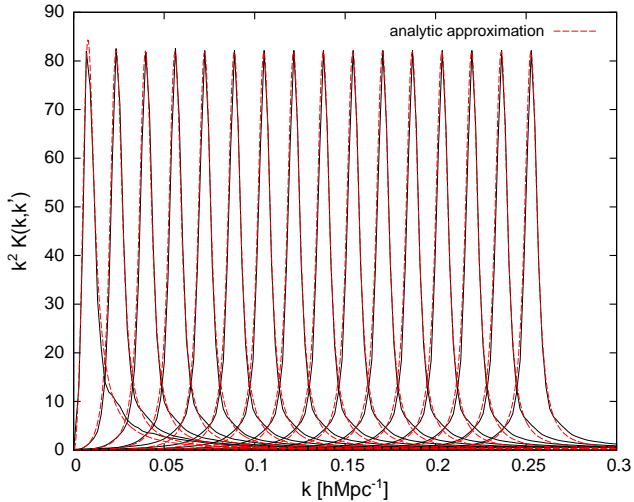}
\caption{Mode coupling kernels. Dashed lines show the analytic approximation given in Eqs. (\ref{eq10})-(\ref{eq11}).}
\label{newfig3}
\end{figure}

Before we can fit model spectra to the data there are several effects one has to take into account:
\begin{itemize}
\item Photometric redshift errors lead to significant damping of the spectrum;
\item Our observed spectrum is a pseudospectrum and thus before fitting one has to convolve model spectra with the survey window;
\item On smaller scales nonlinear effects become noticeable and one needs a model to account for these;
\item And finally, even without photometric redshift errors there are spectral modifications caused by the redshift-space distortions.
\end{itemize}
Our model for the relation between the linear spectrum $P_{\rm{lin}}(k)$ and the observed one $P_{\rm{obs}}(k)$ that tries to accommodate all the four effects listed above reads as:
\begin{equation}\label{eq6}
P_{\rm{obs}}(k) = \int {\rm d}k'\,k'^2P(k')K(k',k)\,,
\end{equation}
where
\begin{equation}\label{eq7}
P(k) = b_{\rm{eff}}^2\left(1+qk^{\frac{3}{2}}\right)f(k)P_{\rm{lin}}(k)\,.
\end{equation}
Here $b_{\rm{eff}}$ is the cluster bias parameter that also incorporates the boost of the power at large scales due to the linear redshift-space distortions. Smaller scale redshift distortions are fully eliminated if the cluster detection algorithm is able to perform a perfect finger-of-God compression. However, in reality the compression is not ideal and thus there should be some extra uncertainties for the central redshift of the galaxy cluster. This can be modelled as an additional smoothing effect in addition to the photometric redshift errors of the individual galaxies. The factor $(1+qk^{3/2})$ is our simplistic model to treat effects due to nonlinear evolution. This form is very similar to the one suggested in \citet{2005MNRAS.362..505C}, however in \citet{2006A&A...459..375H} we found that the power law index $3/2$ provides better fit to the Halo Model\footnote{For a comprehensive review on Halo Model see \citet{2002PhR...372....1C}.} results compared to the value $2$ used in \citet{2005MNRAS.362..505C}. The function $f(k)$ in Eq. (\ref{eq7}) models the damping of the spectrum due to photometric redshift (photo-z) errors. Under the flat sky approximation it is easy to derive the analytic form for $f(k)$ which reads:
\begin{equation}\label{eq8}
f(k)=\frac{\sqrt{\pi}}{2\sigma k}\rm{erf}(\sigma k)\,,
\end{equation}
where the error function $\rm{erf}(x)\equiv\frac{2}{\sqrt{\pi}}\int_0^{x}\exp(-t^2)\,{\rm d}t\,$. This result assumes that photo-z errors follow Gaussian distribution with dispersion $\delta z$ and $\sigma$ is the corresponding spatial smoothing scale, i.e. $\sigma = \frac{c}{H_0}\delta z\,$. And finally, Eq. (\ref{eq6}) models the convolving effect of the survey window. We find that the mode coupling kernels $K(k',k)$ can be well described by the following analytic fit:
\begin{equation}\label{eq09}
K(k',k)=K(k,k')=\frac{C}{kk'}\left[g(k+k')-g(k-k')\right]\,,
\end{equation}
where
\begin{equation}\label{eq10}
g(x)=\arctan\left(\frac{b^4+2a^2x^2}{b^2\sqrt{4a^4-b^4}}\right)\,,
\end{equation}
and the normalization constant $C$ is derived by demanding that $\int K(k,k')k'^2\,\rm{d}k'\equiv 1$, giving:
\begin{equation}\label{eq11}
C=\frac{1}{\pi b \sqrt{2-\left(\frac{b}{a}\right)^2}}\,.
\end{equation}
In the case of the maxBCG survey geometry the best fitting parameters $a$ and $b$ are found to be: $a=0.0044$ and $b=0.0041$. \footnote{A simpler analytic form with $a=b=0.0041$ gives almost as good fit.} For the motivation of this analytic form and for the complete description on how these kernels are determined see \citet{2006A&A...449..891H}. The analytic approximation is compared with the numerically determined coupling kernels in Fig. \ref{newfig3}. We see that our analytic form provides a very good fit to the core parts of the numerical kernels with only slight deviations at the wing portions. In practice, performing the convolution of the model spectra with those approximate kernels in place of the precise numerical ones leads to completely negligible differences.

In this paper we keep the background cosmology fixed to the ``concordance'' flat $\Lambda$CDM model with $\Omega_m=0.27$, $\Omega_b=0.045$, and $h=0.7$. The bias parameters quoted below are measured with respect to the model with $\sigma_8=0.85$. Our decision not to fit for the cosmological parameters is based on two main reasons:
\begin{enumerate}
\item Photo-z uncertainties that are given in \citet{2007astro.ph..1265K} are tested only for the case of the brightest cluster members that have spectroscopic redshifts available. The errors for the whole galaxy population in clusters might differ, and be possibly larger than the numbers given in \citet{2007astro.ph..1265K};
\item Uncertainties in the cluster finding algorithm which are hard to quantify and which might lead to the additional smoothing/overmerging, and thus to the extra uncertainties in derived cluster redshifts.
\end{enumerate}
\begin{figure}
\centering
\includegraphics[width=\plotwd]{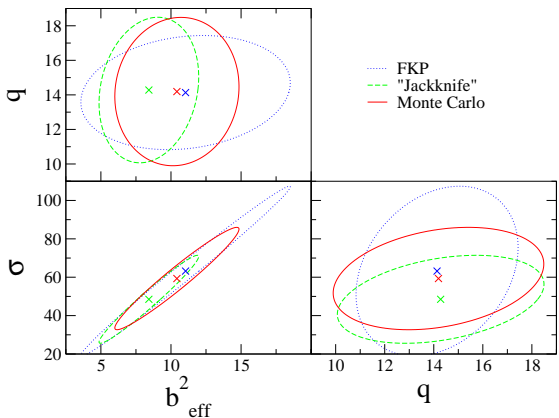}
\caption{Error ellipses for the case with three free parameters. The solid, dashed, and dotted lines correspond to the Monte Carlo, ``jackknife'' and FKP error models, respectively. The crosses mark the best fitting parameter values. Note the strong degeneracy between $b_{\rm{eff}}^2$ and $\sigma$.}
\label{fig}
\end{figure}
Any untreated systematics in the determination of the radial smoothing scale $\sigma$ will start to interfere with the cosmological parameter estimation, resulting in biased estimates. Due to the uncertainties listed above we first treat the radial smoothing scale $\sigma$ as a free parameter. Thus we are left with three free parameters: $b_{\rm{eff}}$, $\sigma$, $q$. We also investigate separately the case where the radial smoothing scale is fixed at $c/H_0 \cdot \delta z \simeq 30\,h^{-1}\,\rm{Mpc}$ as suggested in \citet{2007astro.ph..1265K}. Some of the results of these calculations are presented in Figs. \ref{fig4} and \ref{fig5}, and in Tables \ref{tab1} and \ref{tab2}.  
\begin{table*}
\caption{Best fitting model parameters and $\chi^2$ values along with the inferred confidence levels for BAO detection for the case with three free parameters: $b_{\rm{eff}}$, $q$, $\sigma$.}
\label{tab1}
\begin{tabular}{|l|c|c|c|c|c|c|}
\hline
 $P(k)$ & \multicolumn{6}{c|}{3 free parameters: $b_{\mathrm{eff}}^2$, $q$, $\sigma$; $16-3=13$ dof; expected $\chi^2\simeq 13.0 \pm 5.1$}\\
 \cline{2-7}
 error model & $b_{\mathrm{eff}}^2$ & $q$ & $\sigma$ & $\chi_{\mathrm{wiggly}}^2$ & $\chi_{\mathrm{smooth}}^2$ & BAO detection \\
\hline
FKP & $11.0 \pm 4.9$ & $14.1 \pm 2.2$ & $63 \pm 29$ & $7.42$ & $11.0$ & $1.9\sigma$ \\
``Jackknife'' & $8.4 \pm 2.3$ & $14.3 \pm 2.8$ & $49 \pm 15$ & $9.57$ & $14.5$ & $2.2\sigma$\\
Monte Carlo & $10.4 \pm 2.9$ & $14.2 \pm 2.8$ & $59 \pm 18$ & $9.41$ & $14.1$ & $2.2\sigma$\\
\hline
\end{tabular}
\end{table*}
\begin{table*}
\caption{Analog of Table \ref{tab1} for the case with two free parameters: $b_{\rm{eff}}$, $q$, and the value for $\sigma$ is fixed to $30\,h^{-1}\,\rm{Mpc}$.}
\label{tab2}
\begin{tabular}{|l|c|c|c|c|c|}
\hline
 $P(k)$ & \multicolumn{5}{c|}{$\sigma=30$; 2 free parameters: $b_{\mathrm{eff}}^2$, $q$; $16-2=14$ dof; expected $\chi^2\simeq 14.0 \pm 5.3$}\\
 \cline{2-6}
 error model & $b_{\mathrm{eff}}^2$ & $q$ & $\chi_{\mathrm{wiggly}}^2$ & $\chi_{\mathrm{smooth}}^2$ & BAO detection \\
\hline
FKP & $5.64 \pm 0.32$ & $11.9 \pm 1.9$ & $11.5$ & $14.3$ & $1.7\sigma$ \\
``Jackknife'' & $5.70 \pm 0.37$ & $12.1 \pm 2.4$ & $12.4$ & $15.7$ & $1.8\sigma$\\
Monte Carlo & $5.75 \pm 0.43$ & $11.5 \pm 2.5$ & $15.2$ & $19.4$ & $2.0\sigma$\\
\hline
\end{tabular}
\end{table*}
To calculate theoretical model spectra we have used fitting formulae for the transfer functions presented in \citet{1998ApJ...496..605E}. In the lower panels of Figs. \ref{fig4} and \ref{fig5} we have divided the spectra by the ``smooth'' model spectrum with acoustic oscillations removed (see \citealt{1998ApJ...496..605E} for details). For fitting purposes we have used Levenberg-Marquardt  $\chi^2$-minimization technique as described in \citet{1992nrfa.book.....P} with the additional modification to allow for the nondiagonal data covariance matrices. We perform the fitting for all three error models: (i) FKP errors with diagonal covariance matrix, (ii) ``jackknife'' and (iii) Monte Carlo errors along with full covariance matrices. The solid lines in Figs. \ref{fig4} and \ref{fig5} represent the best fitting ``wiggly'' models for the case of Monte Carlo errors. In these figures we also demonstrate the damping effect of the window convolution. The corresponding linear spectra are shown with short dashed lines. One can see that the shape of the measured spectrum deviates strongly from the linear model spectrum, which is mainly driven by the photo-z errors. The best fitting parameter values along with the corresponding $\chi^2$ values are given in Tables \ref{tab1}, \ref{tab2}. There we also present the $\chi^2$ value for the best fitting ``smooth'' model. In general we can see that the results for the $\chi^2$-statistic agree well with expectations $13.0 \pm 5.1$ and $14.0 \pm 5.3$ for $13$ and $14$ degrees of freedom, relevant for Table \ref{tab1} and \ref{tab2}, respectively. Comparing the $\chi^2$ values for the best fitting ``smooth'' and ``wiggly'' models one can assess the confidence level for the BAO detection, the results of which are presented in the last columns of Table \ref{tab1} and \ref{tab2}. We note that using the full ``jackknife'' and and Monte Carlo covariance matrices in place of a simple FKP error model we obtain higher confidence for the detection of the oscillatory features in the spectrum. This is easy to understand: our first error model, where we assume diagonal covariance matrix and FKP errors is certainly not realistic. Survey window, and more importantly nonlinear mode-mode coupling induces significant correlations between neighboring power spectrum bins. Stronger correlations amongst the bins make the spectrum ``harder to distort'', and thus any surviving oscillatory features gain higher statistical weight. It is interesting to note that in case we allow $\sigma$ to be a free parameter the best fitting values are certainly larger than $30\,h^{-1}\,\rm{Mpc}$. Also the bias parameter $b_{\rm{eff}}$ has a noticeable increase compared to the $\sigma = 30\,h^{-1}\,\rm{Mpc}$ case. In fact there is a strong degeneracy between $b_{\rm{eff}}$ and $\sigma$: if one chooses to increase $b_{\rm{eff}}$, one can still obtain a good fit to the data by also increasing $\sigma$ by a suitable amount (see Fig. \ref{fig}). The fact that the case with three free parameters favors higher value for $\sigma$ (and thus also for $b_{\rm{eff}}$) is mainly driven by the relatively large amplitudes of the two first power spectrum bins. On the other hand, if one assumes that $\sigma = 30\,h^{-1}\,\rm{Mpc}$ is a reasonably correct value then one is left with some discrepancy on the largest scales between the model and data (see Fig. \ref{fig5}). It is interesting that some large-scale ``extra power'' has also been detected by several other authors e.g. \citet{2006astro.ph..5302P,2007MNRAS.374.1527B} who analyzed SDSS photometric LRG sample.\footnote{See also Fig. 4 in \citealt{2006A&A...449..891H} where SDSS spectroscopic LRG sample reveals some mild excess of power on the largest scales.} Of course these mild discrepancies on the largest scales might be due to inaccurate treatment of the survey selection effects. Taking into account several uncertainties concerning the appropriate value for $\sigma$, and also uncertainties concerning the survey selection effects, we are here unfortunately unable to draw any firm conclusions about the existence of the excess large-scale power. This issue certainly deserves a dedicated study and currently the best available sample for this purpose is probably the SDSS LRG spectroscopic sample.

\begin{figure*}
\centering
\includegraphics[width=\plotwdtwo]{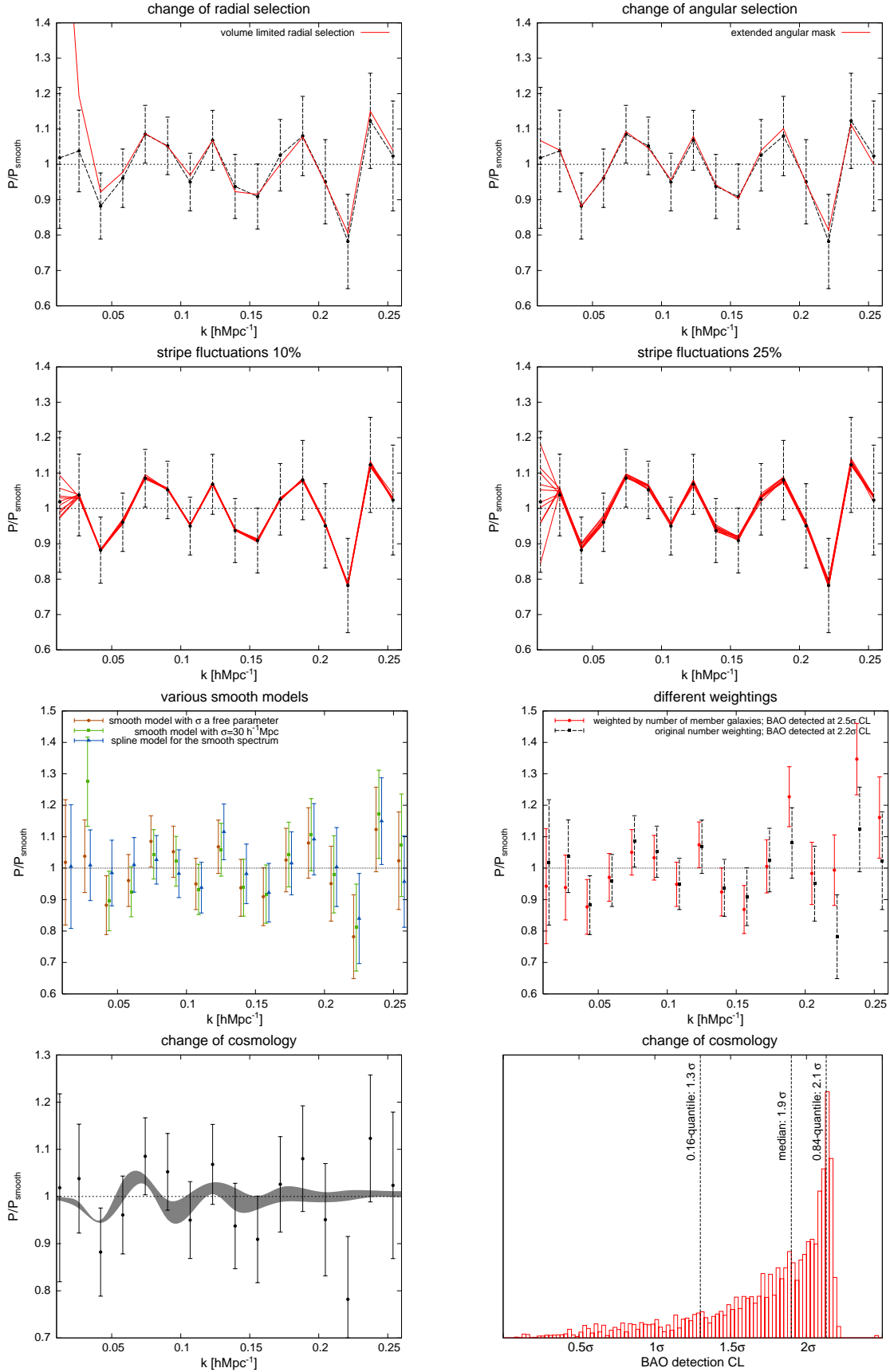}
\caption{Results of several stability tests. See the main text for a detailed explanation.}
\label{fig8}
\end{figure*}

However, one extra exercise we can perform here is to try to investigate how well the obtained values for the bias parameter $b_{\rm{eff}}$ agree with the model expectations. If one assumes a flat $\Lambda$CDM model with the parameters as given above (i.e. $\Omega_m=0.27$, $\sigma_8=0.85$) then for the mass-limited cluster survey covering the same redshift range and sky area as maxBCG sample, and giving the same number of clusters, one infers a lower mass limit of $M_{\rm{low}}=7.1\times10^{13}\,h^{-1}M_{\odot}$. We have calculated $M_{\rm{low}}$ using the Sheth-Tormen mass function \citep{1999MNRAS.308..119S} averaged over the past light-cone (see \citealt{2006A&A...446...43H} for details). The corresponding light-cone averaged effective bias parameter including increase of power due to large-scale redshift-space distortions (see again \citealt{2006A&A...446...43H}) with respect to the $z=0$ linear spectrum turns out to be $b_{\rm{eff}}\simeq 2.8$ ($b_{\rm{eff}}^2 \simeq 7.8$).\footnote{In reality of course for the mass-limited sample the number density drops with redshift, while maxBCG catalog is close to being volume-limited. We have calculated that to achieve a comoving number density of $3.2 \times 10^{-5} \,h^{3}\,\rm{Mpc}^{-3}$, as relevant for the maxBCG catalog, the lower mass limits for redshifts $0.1$ and $0.3$ should be $7.8\times10^{13}\,h^{-1}M_{\odot}$ and $6.6\times10^{13}\,h^{-1}M_{\odot}$, respectively. The effective bias parameters with respect to the $z=0$ linear matter power spectrum for these redshifts and lower mass limits are $b_{\rm{eff}}^{z=0.1}\simeq 2.75$ and $b_{\rm{eff}}^{z=0.3}\simeq 2.78$, i.e. basically independent of redshift, and also agreeing very well with the light-cone averaged value of $2.8$ quoted in the main text. This approximate constancy of the clustering amplitude at different redshifts for the objects with fixed comoving number density is a well known result.} Comparing this value of $b_{\rm{eff}}^2\simeq 8$ with the ones given in Tables \ref{tab1} and \ref{tab2} one sees that $\sigma = 30\,h^{-1}\,\rm{Mpc}$ case has somewhat lower bias value, whereas the case with free $\sigma$ tends to have values for $b_{\rm{eff}}^2$, which seem to agree reasonably well (especially true for the case with ``jackknife'' errors). Thus this simple analysis would favor the ``extra smoothing'' hypothesis over the ``extra power'' one.

\section{Stability of the results}\label{sec4}
In this Section we are going to relax several assumptions that were made in our original analysis in order to test for the stability of the BAO features. The results of these investigations are presented in Fig. \ref{fig8}. We start by testing the stability of the BAO against possible uncertainties in the survey selection function. For this purpose we have repeated power spectrum calculations with modified radial and angular selections. On the upper left-hand panel of Fig. \ref{fig8} we show the change induced in the BAO if one replaces the smooth cubic spline fitted radial selection with the one corresponding to the volume-limited sample (see also Fig. \ref{fig3}). Here the solid line represents the case with a new radial selection whereas the dashed line with errorbars corresponds to our original BAO measurement.\footnote{Here and in the following we are going to show only FKP errors} We see that at almost all the scales, except for the very largest ones, the BAO features are essentially unchanged. However, this exercise shows that the largest scales are very vulnerable with respect to the unaccounted slow trends in the survey selection. The results of the similar exercise for the angular selection function are shown on the upper right-hand panel of Fig. \ref{fig8}. Here the agreement between the BAO measurements using the ``minimal'' and ``maximal'' angular masks (see Section \ref{sec2} for explanation) is almost ideal. Thus we can conclude that our results seem to be robust with respect to the uncertainties in the survey selection, which is especially true for scales with wavenumbers $k\gtrsim 0.03 \,h\,\rm{Mpc}^{-1}$.

The uncertainties in the SDSS photometry might lead to the variations in the completeness across the survey area, and thus might potentially influence the BAO features. The SDSS imaging data is collected along $2.5^{o}$ wide scan stripes. In order to test the effect of varying completeness we have created several random catalogues where the completeness (and thus the number density of random points) has been allowed to fluctuate from stripe to stripe. On the second left-hand panel of Fig. \ref{fig8} we have shown the results of this analysis for the case where the completeness was allowed to vary uniformly randomly in the range $0.9-1.0$. The corresponding right-hand panel shows the case where the fluctuations were drawn from the interval $0.75-1.0$, which should be a very conservative choice keeping in mind a superb photometric quality of the SDSS data. In both cases we have plotted ten realizations just to give an idea of the possible induced scatter. It is clear from these investigations that possible stripe to stripe completeness fluctuations have negligible effect on the extracted BAO features. The only noticeable variations occur only on the very largest scales, $k\lesssim 0.03 \,h\,\rm{Mpc}^{-1}$.  

The third topmost panel on the left-hand side shows the effect of changing smooth models with respect to which the BAO are defined. There we have shown the cases were the smooth model is defined by Eqs. (\ref{eq6}) and (\ref{eq7}) with the ``no-wiggle'' form of the linear transfer function given as in \citet{1998ApJ...496..605E}. We have plotted the results when the smoothing scale $\sigma$ was taken to be a free parameter and also when it was fixed to the value of $30\,h^{-1}\,\rm{Mpc}$. The third smooth model was defined as a cubic spline fit to the measured power spectrum with the spline nodes starting at wavenumber $0.025 \,h\,\rm{Mpc}^{-1}$ with a uniform spacing of $0.05 \,h\,\rm{Mpc}^{-1}$.\footnote{For this choice of nodes see e.g. \citealt{2007ApJ...657...51P}.} As it was probably expected, with the low accuracy of the current BAO measurement the results are rather insensitive to how one precisely defines a smooth model. 

In our power spectrum calculations we have so far used FKP weighting, which is a function of distance only. However, with galaxy clusters, in addition to FKP weight, we can also apply an extra weight that is proportional to the number of member galaxies. This way we give more statistical weight to massive systems, and thus one expects to measure higher clustering strength. In addition one might expect that photometric redshifts are more accurate for the systems containing larger number of member galaxies. This is indeed what we find: the large-scale clustering amplitude is slightly higher and the small-scale damping is mildly weaker compared to the case with the original weights only. Larger clustering amplitude along with smaller damping leads to tighter errorbars on power spectrum measurement. This can be seen on the third right-hand panel of Fig. \ref{fig8}. Compared to the initial weighting the confidence level for the BAO detection is increased from $2.2 \sigma$ to $2.5 \sigma$. Apart from slightly different confidence for the BAO detection both spectra have oscillatory features that are in good agreement with each other.

The lower panels in Fig. \ref{fig8} show the effect of changing the fiducial cosmological model. For this purpose we have generated a MCMC chain with $25,000$ elements\footnote{We have used CosmoMC package \citep{2002PhRvD..66j3511L} available at http://cosmologist.info/cosmomc/} for flat $\Lambda$CDM models using WMAP5\footnote{http://map.gsfc.nasa.gov/} data \citep{2008arXiv0803.0593N} For each of these cosmologies spectra in the parametric form of Eqs. (\ref{eq6}) and (\ref{eq7}) are fitted to the data. Also, as the original power spectrum measurements were done by adopting a distance-redshift relation of one particular fiducial cosmology, all of the model spectra are transformed to the ``fiducial model framework'' in a manner described in \citet{2006A&A...459..375H}. The gray ``snake'' on the lower left-hand panel shows the $1\sigma$ region filled with the best fitting models. The corresponding right-hand panel shows the histogram of the BAO detection confidence level for all of the cosmologies tested. As one can see, for a typical flat $\Lambda$CDM model consistent with the WMAP5 data BAO is detected at $\sim 2\sigma$ confidence level. 

There are other possible effects that might influence the BAO, e.g. the performance of the cluster finder might vary with angular position and distance and thus lead to fluctuations in completeness and purity of the final cluster sample. However, it  is hard to imagine why cluster detector should work in a spatially correlated way and imprint a $\sim 100 \,h^{-1}\,\rm{Mpc}$ preferable scale to the clustering pattern. Thus it is unlikely that the BAO will be noticeably affected by those means: even if cluster finding efficiency and fraction of "false systems" changes with redshift these changes are gradual i.e. can only lead to the changes in the broad-band shape of the spectrum. 

In addition, there are always worries related to biasing, i.e. how the fluctuating density field as inferred from the spatial distribution of the tracer objects is related to the underlying matter distribution. Here we would like to argue that biasing is probably not a big issue, e.g. if one takes N-body simulations of the ``concordance'' cosmology then the typical displacements of the particles are $\sim 10 \,h^{-1}\,\rm{Mpc}$. These are an order of magnitude smaller than the acoustic scale, i.e. in that respect one can look at the structure formation as a fairly local process. These small displacements can lead only to smooth distortions but cannot produce sharp ringing features in the spectrum.

As an additional test we have investigated the influence of the assumed photo-z error model on the detected BAO signal. So far in this paper we have used a simple model where the photo-z errors are independent of redshift, whereas in \citet{2007astro.ph..1265K} the authors find a mild evolution where the errors become slightly larger as the distance increases. We have used a simple model where the errors increase linearly with redshift $\delta z (z)= (z - 0.1)/(0.3 - 0.1)0.011 + (0.3 - z)/(0.3 - 0.1)0.006$, which gives a good fit to the trend observed in \citet{2007astro.ph..1265K}. It turns out that our results in this case are essentially indistinguishable from the simpler model assumed earlier. Having a model with redshift dependent photo-z errors our model spectra are now calculated as light-cone averages following the description in \citet{2006A&A...446...43H} (see also \citet{1999ApJ...527..488Y}). Thus the model spectra are now sums over spectra with different damping factors. However, for the spatial smoothing scales $\sigma \sim 30\,h^{-1}\,\rm{Mpc}$ the error function in Eq. (\ref{eq8}) can be simply set to one for wavenumbers $k\gtrsim 0.07 \,h\,\rm{Mpc}^{-1}$ (with an accuracy better than $0.5\%$), and so our light-cone spectra are effectively sums over spectra which at most relevant $k$ values have approximate power law form, with the only difference in their amplitude. Since the sum over power laws with varying amplitude and the same spectral index is still a power law this should explain why we do not observe any significant deviation from the case when only a single effective smoothing scale is used for all the redshifts. Also, the small biases of the photo-z errors as measured in \citet{2007astro.ph..1265K} turn out to have negligible effect on our final results. What concerns the maxBCG clusters then actually we are quite lucky as the availability of spectroscopic redshifts for many of the galaxies allows one to test the reliability of the photo-z errors, and indeed, as seen in Fig. 9 of \citet{2007astro.ph..1265K} the Gaussian photo-z error model seems to be quite adequate.

In general, for the photo-z errors to have a strong impact on the BAO demands a nontrivial correlation structure over scales of $\sim 100\,h^{-1}\,\rm{Mpc}$. This might happen at some specific redshifts, but it is quite unrealistic for the errors to be correlated at those scales throughout the full survey volume.

As a final test we have investigated how the significance of the results change for different binnings of the power spectrum. If we use two times smaller wavenumber binning (i.e. instead of 16 bins we have now 32 bins) the confidence levels for the BAO detection given in Table \ref{tab1} get replaced by $1.8\sigma$, $2.2\sigma$, and $2.0\sigma$ for the FKP, ``jackknife'', and Monte Carlo error models, respectively. The new values for Table \ref{tab2} are: $1.6\sigma$, $1.8\sigma$, and $1.8\sigma$.
 
\begin{figure}
\centering
\includegraphics[width=\plotwd]{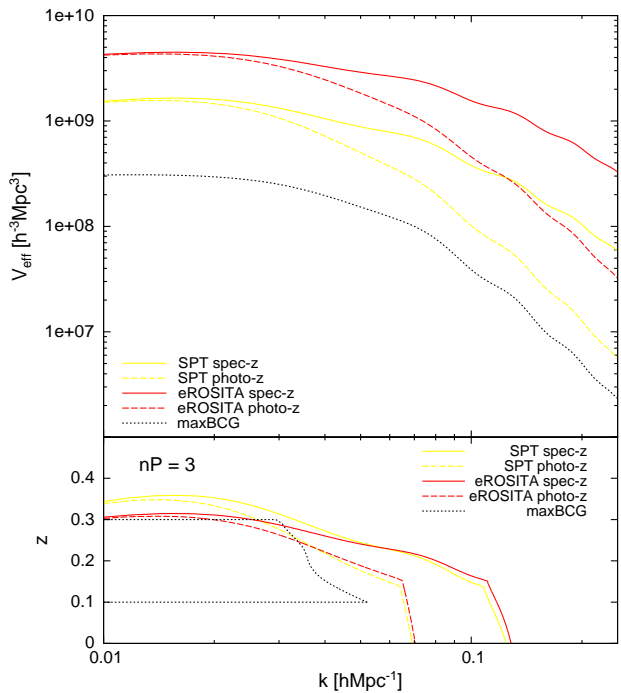}
\caption{Upper panel: Comparison of the maxBCG effective volume with the SPT-like SZ survey covering one octant of the sky with the total of $\sim 20,000$ galaxy clusters, and with the eROSITA-type X-ray survey over $50\%$ of the sky yielding $\sim 100,000$ clusters. Here the two scenarios have been plotted: (i) the optimistic case where one has the spectroscopic redshifts for all the clusters, (ii) the case with the photo-z errors comparable to the accuracy obtained for the maxBCG clusters, i.e. $\delta z \simeq 0.01$. Lower panel: Contours $nP=3$. Below those lines the sampling density is high enough for the shot noise not to contribute significantly. The curve for the maxBCG sample is restricted to a redshift range $z=0.1-0.3$.}
\label{newfig4}
\end{figure}

\section{Conclusions and discussion}
We have calculated the redshift-space power spectrum of the maxBCG cluster sample \citep{2007astro.ph..1265K} -- currently the largest cluster catalog in existence. After correcting for the radial smoothing caused by the photometric redshift errors, treating the convolution with a survey window, and introducing a simple model for the nonlinear effects we show that ``concordance'' $\Lambda$CDM model is capable of providing very good fit to the data. Moreover, we are able to find weak evidence for the acoustic features in the spatial clustering pattern of the maxBCG sample. If radial smoothing scale $\sigma$, effective bias parameter $b_{\rm{eff}}$, and nonlinear distortion parameter $q$ are treated as completely free parameters the model with acoustic oscillations is favored by $2.2\sigma$ over the corresponding ``smooth'' model without any oscillatory behavior. The evidence for BAO weakens somewhat ($2.0\sigma$ in case of Monte Carlo errors), and we are also left with some ``extra power'' on large scales, if we fix the smoothing scale $\sigma$ to $30\,h^{-1}\,\rm{Mpc}$ as suggested by the photometric redshift errors $\delta z = 0.01$ estimated in \citet{2007astro.ph..1265K}. We note that these photometric errors actually apply only for the very brightest cluster members and in reality the scatter for the whole sample might be larger. In addition the cluster finding algorithm itself might introduce extra smoothing/overmerging along the line of sight. Due to these uncertainties we have decided not to carry out any cosmological parameter study in this paper, as any untreated systematics will be immediately propagated to the parameter estimates, leading to biased results. 

We have also carried out power spectrum measurement by applying additional weights that are proportional to the richness of clusters. This way we give more statistical weight to systems with a higher mass. As a result we find slightly higher large-scale clustering amplitude and mildly weaker small-scale damping of the spectrum. The latter effect is probably caused by tighter photometric redshift errors achievable for richer clusters. Larger clustering amplitude along with smaller damping leads to tighter errorbars on power spectrum measurement. Compared to the initial weighting the confidence level for the BAO detection is increased from $2.2 \sigma$ to $2.5 \sigma$.\footnote{This corresponds to the case where the spatial smoothing scale due to photometric redshift errors was treated as a free parameter.} 

In our paper we have estimated power spectrum errors using three different methods: (i) FKP errors, (ii) ``jackknife'' errors, obtained by dividing the survey into $5\times 5 \times 3$ chunks containing equal number of galaxies, (iii) Monte Carlo errors found from 1000 mock catalogs. All of these errors turn out to agree very well with each other. We have tested the stability of our power spectrum measurements by relaxing various assumptions made during the main part of our analysis. The results of this study, which were presented in Section \ref{sec4}, demonstrate the reliability of the original analysis.

The detectability of BAO with $\sim 10^4$ galaxy clusters is very encouraging result keeping in mind the future cluster surveys, such as the ones based on the measurement of the thermal Sunyaev-Zeldovich (SZ) effect \citep{1972CoASP...4..173S}, e.g. {\sc Spt}\footnote{http://spt.uchicago.edu/}, {\sc Planck}\footnote{www.rssd.esa.int/Planck/}, or the proposed $100,000$-cluster X-ray survey eROSITA\footnote{www.mpe.mpg.de/erosita/MDD-6.pdf}. On the upper panel of Fig. \ref{newfig4} we have compared the effective volume of the maxBCG cluster sample with the SPT-like SZ survey covering one octant of the sky with the total of $\sim 20,000$ galaxy clusters, and with the eROSITA-type X-ray survey over $50\%$ of the sky yielding $\sim 100,000$ clusters. We have plotted two distinct scenarios: (i) the optimistic case where one has the spectroscopic redshifts for all the clusters, (ii) the case with the photo-z errors comparable to the accuracy obtained for the maxBCG clusters, i.e. $\delta z \simeq 0.01$. Thus, the eROSITA-like cluster survey with spectroscopic cluster redshifts has approximately $50$ times larger effective volume compared to the maxBCG sample at $k \sim 0.1 \,h\,\rm{Mpc}^{-1}$ implying $\sim 7$ times smaller errors for $P(k)$. If only photometric redshifts with  $\delta z \simeq 0.01$ are available the effective volume increases by a factor of $\sim 12$ (which is also the case for the SPT-like survey with spectroscopic redshifts) leading to $\sim 3.5$ times decrease in power spectrum errors. The lower panel of Fig. \ref{newfig4} shows the contours on the wavenumber--redshift plane where $nP=3$, which is a typical target value of signal to noise for the future BAO surveys involving galaxies (see e.g. \citealt{2006astro.ph..9591A}.) Below those lines the sampling density is high enough for the shot noise not to contribute significantly. The curve for the maxBCG sample is restricted to a redshift range $z=0.1-0.3$. It is evident that except for the lowest redshifts, i.e. $z \lesssim 0.3$, those future cluster surveys will have significant levels of discreteness noise, and thus in terms of the raw performance of measuring the BAO only can not compete with a dedicated galaxy surveys.

In reality the main target of these cluster surveys is to map the evolution of the cluster number density as a function of redshift, since this quantity is very sensitive to the amplitude and to the growth rate of the density perturbations, and as such it provides a powerful tool to constrain the properties of DE. The possibility to also find the clustering power spectrum can be seen as a useful byproduct of these surveys that comes essentially ``for free''. However, both cluster number count study and spatial clustering analysis needs as an input estimates for the cluster redshifts, with the former probably doing relatively well with rather poor redshift estimates (e.g. photometric redshifts). This is a rather complicated task since one arguably needs of order $10$ galaxy redshifts per cluster to reliably infer cluster redshift. In that respect for the measurement of BAO it is certainly less costly to sample the cosmic density field using e.g. LRGs or blue emission line galaxies. The latter type of objects are targets for the currently ongoing WiggleZ project \citep{2007astro.ph..1876G} at AAO that attempts to measure redshifts for $400,000$ objects over $\sim 1000$ deg$^2$ and cover the redshift range $0.5< z < 1$. There is also a {\sc Wfmos} instrument (see e.g. \citealt{2005A&G....46e..26B}) construction planned for the Gemini and Subaru observatories that will hopefully be completed in 2012. This new multi-object spectrograph will be able to measure the spectra of $\sim 5000$ objects at the same time. There are plans to perform a wide field ($\sim 2000$ deg$^2$) redshift survey giving spectra for $\sim 2 \times 10^6$ galaxies up to redshifts of $z \sim 1.3$ together with a narrower ($\sim 200$ deg$^2$) and deeper ($z \sim 2 - 3$) survey with a yield of $\sim 5 \times 10^5$ galaxies. For the more distant future (around 2020) one would expect a superb measurement of the matter power spectrum using $\sim 10^9$ galaxy redshifts obtainable via the 21 cm measurements by the Square Kilometre Array ({\sc Ska})\footnote{http://www.skatelescope.org/} (e.g. \citealt{2004NewAR..48.1063B,2005MNRAS.360...27A}). However, before {\sc Wfmos} and {\sc Ska} become available several wide area imaging surveys, such as Pan-STARRS\footnote{pan-starrs.ifa.hawaii.edu/} or Dark Energy Survey\footnote{https://www.darkenergysurvey.org/} ({\sc Des}), will be performed. Although these photometric surveys lack accurate redshift information, the huge number of objects detectable over large sky areas significantly compensate this shortcoming, allowing us to obtain a highly competitive measurement of BAO \citep{2005MNRAS.363.1329B}.

NOTE: \\ Several months after the first version of this work was submitted a paper appeared \citep{2009ApJ...692..265E} where the authors perform correlation function analysis using the same maxBCG cluster sample. They also find weak evidence for the BAO, however their confidence level for the BAO detection ($1.4- 1.7\, \sigma$) is somewhat lower than we typically find in our study ($1.7-2.2 \sigma$). It is important to point out that similar to our work \citet{2009ApJ...692..265E} consider only isotropised quantities. Due to photo-z errors the real clustering pattern is significantly anisotropic and it is clear that isotropization looses quite some information. Here one should realize that in general the information loss for the case of the power spectrum somewhat differs from the information loss in the case of the correlation function, or equivalently, isotropised power spectrum and correlation function do not form a Fourier transform pair (only the full anisotropic quantities are the Fourier transforms of each other). Taking this into account one should not \emph{a priori} expect similar results for the BAO detection.

It was somewhat surprising to find out that \citet{2009ApJ...692..265E} use only five times more random points to build their unclustered catalogs. As the cluster sample itself is very sparse this surely leads to a very noisy reference level with respect to what the fluctuations are measured, leaving serious doubts on the convergence of their results.

\citet{2009ApJ...692..265E} estimate the correlation function covariance matrix in two different ways: (i) ``jackknife'' method, (ii) simple transformation of the power spectrum covariance. In their ``jackknife'' method the subsamples are generated by each time randomly removing $1/1000$th of the clusters. This is a valid method only as long as the clusters are independent. However, due to the large-scale structure the clusters are not independent, and thus one should use methods which are able to cope better with the situation in hand. One method of dealing with dependent data is to replace the usual bootstrap/jackknife method with the block bootstrap/jackknife scheme (see e.g. \citealt{lahiri.book}). This is indeed the reason why in large-scale structure studies the ``jackknife'' method is usually implemented by removing whole contiguous spatial chunks of data, instead of just some random set of points. By removing whole spatial regions one keeps the correlations inside the regions. If one removes random selection of points those correlations are lost, which leads to an underestimation of the off-diagonal elements in the power spectrum and correlation function covariance matrices. This might explain why their ``jackknife'' error model gives weaker confidence level for the BAO detection compared to our case. The other method of estimating covariance matrix is actually equivalent to our diagonal FKP covariance matrix case. Indeed, their Eq. (17) assumes that power spectrum covariance matrix is diagonal. \footnote {In case of the non-diagonal power spectrum covariance matrix this equation should have a double integral over the wavenumbers, instead.} Looking at their Table 4 the relevant entry $\Delta \chi^2=2.4$ is quite similar to what we find, $\Delta \chi^2=2.8$ (see Table \ref{tab2}), and thus is in a reasonable agreement with our FKP covariance matrix case.

\section*{Acknowledgments}
I thank Ofer Lahav and the Referee for useful comments and suggestions.
I acknowledge the support provided through a PPARC/STFC postdoctoral fellowship at UCL and travel support from ETF grant 7146.

Funding for the SDSS and SDSS-II has been provided by the Alfred P. Sloan Foundation, the Participating Institutions, the National Science Foundation, the U.S. Department of Energy, the National Aeronautics and Space Administration, the Japanese Monbukagakusho, the Max Planck Society, and the Higher Education Funding Council for England. The SDSS Web Site is http://www.sdss.org/.

The SDSS is managed by the Astrophysical Research Consortium for the Participating Institutions. The Participating Institutions are the American Museum of Natural History, Astrophysical Institute Potsdam, University of Basel, University of Cambridge, Case Western Reserve University, University of Chicago, Drexel University, Fermilab, the Institute for Advanced Study, the Japan Participation Group, Johns Hopkins University, the Joint Institute for Nuclear Astrophysics, the Kavli Institute for Particle Astrophysics and Cosmology, the Korean Scientist Group, the Chinese Academy of Sciences (LAMOST), Los Alamos National Laboratory, the Max-Planck-Institute for Astronomy (MPIA), the Max-Planck-Institute for Astrophysics (MPA), New Mexico State University, Ohio State University, University of Pittsburgh, University of Portsmouth, Princeton University, the United States Naval Observatory, and the University of Washington.

\bibliographystyle{mn2e}
\bibliography{aamnem99,references}

\bsp

\label{lastpage}

\end{document}